\documentclass[tradiabstract]{aa}            % for a print version

\usepackage{natbib}
\bibpunct{(}{)}{;}{a}{}{,}
\usepackage{graphicx}
\usepackage{txfonts}

\begin{document}

\title{An analysis of the transit times of CoRoT-1b\thanks{Based on data obtained with the CoRoT satellite, which was developed and is operated by the CNES, with participation of the Science Program of ESA, ESTEC/RSSD, Austria, Belgium, Brazil, ESA, Germany, and Spain.}}

\author{J.~L. Bean\inst{1}}

\institute{Institut f\"{u}r Astrophysik, Georg-August-Universit\"{a}t G\"{o}ttingen, Friedrich-Hund-Platz 1, 37077 G\"{o}ttingen, Germany\\
\email{bean@astro.physik.uni-goettingen.de}}

\date{Accepted for publication in A\&A}

\abstract{The CoRoT satellite is expected to discover tens of new transiting exoplanets during its mission. For each of these planets there will be a resulting long, continuous sequence of transit times that can be used to search for perturbations arising from an additional planet in the system. I report the results from a study of the transit times for CoRoT-1b, which was one of the first planets discovered by CoRoT. Analysis of the pipeline reduced CoRoT light curve yields a new determination of the physical and orbital parameters of planet and star, along with 35 individual transit times at a typical precision of 36\,s. I estimate a planet-to-star radii ratio of $R_{p}/R_{\star}\,=\,0.1433\,\pm\,0.0010$, a ratio of the planet's orbital semimajor axis to the host star radius of $a/R_{\star}\,=\,4.751\,\pm\,0.045$, and an orbital inclination for the planet of $i=83.88\degr\pm\,0.29\degr$. The observed transit times are consistent with CoRoT-1b having a constant period and there is no evidence of an additional planet in the system. I use the observed constancy of the transit times to set limits on the mass of a hypothetical additional planet in a nearby, stable orbit. I ascertain that the most stringent limits (4\,M$_{\oplus}$ at 3\,$\sigma$ confidence) can be placed on planets residing in a 1:2 mean motion resonance with the transiting planet. In contrast, the data yield less stringent limits on planets near a 1:3 mean motion resonance (5\,M$_{Jup}$ at 3\,$\sigma$ confidence) than in the surrounding parameter space. In addition, I use a simulation to investigate what sensitivity to additional planets could be obtained from the analysis of data measured for a similar system during a CoRoT long run (100 sequential transit times). I find that for such a scenario, planets with masses greater than twice that of Mars (0.2\,M$_{\oplus}$) in the 1:2 mean motion resonance would cause high-significance transit time deviations. Therefore, such planets could be detected or ruled out using CoRoT long run data. I conclude that CoRoT data will indeed be very useful for searching for planets with the transit timing method.}


\maketitle

\section{Introduction}
Transiting exoplanets present unique opportunities for observational study. It is only for these planets that masses, radii, and atmospheric properties together may be determined. These data give important insight into the planets' structures, and thus, their formation and evolutionary history as well. Additionally, observations of transiting planets allow very precise characterization of their orbits, and this can be leveraged to investigate the architecture of the systems the planets reside in. The idea behind this is that another planet in the same system will perturb the transiting planet's orbit. These orbital variations could be detectable in transit observations over time, and they could be used to characterize the perturbing planet without the need to detect it using additional observational methods (e.g. radial velocities). On the other hand, the absence of such observed variations can be used to place limits on the mass and orbital properties of a hypothetical additional planet. 

The main method used to look for perturbations arising from an additional planet in a transiting planet system is the search for transit timing variations \citep[TTVs,][]{agol05, holman05}. The premise is that perturbations to a transiting planet's orbit from another planet will cause deviations from strict periodicity. By regularly measuring the time of central transit for the transiting planet one would observe increasing deviation from the expected constant ephemeris. For example, a typical data set of measured transit times with precisions on the order of a few tens of seconds and spread out over a few years can be sensitive to perturbations from additional planets with masses of 1\,M$_{\oplus}$ or even lower in or near low-order mean motion resonances with the transiting planet. Therefore, TTV investigations can probe for planets in a unique region of parameter space.

To date, there have been no definitive detections of transit timing variations attributable to the existence of an additional planet in a transiting planet system, although there are some systems like OGLE-TR-111 \citep{diaz08} that do warrant further observation and study to determine whether some limited discrepant data are indicative of true TTVs. As a result of no definitive detected variations, most work in this area has been on establishing baselines for long-term monitoring \citep[e.g. the Transit Light Curve project,][]{holman06} and setting limits on the existence of additional planets using the observed constancy as a constraint.

The transiting planet systems for which detailed calculations have been carried out to place limits on additional planets based on no observed TTV variations are TrES-1 \citep{steffen05}, HD\,209458 \citep{agol07, miller-ricci08a}, HD\,189733 \citep{miller-ricci08b}, and GJ\,436 \citep{bean08}. The upper limits to additional planets in these systems are interesting for a variety of reasons. For the three gas giant transiting planets (TrES-1b, HD\,209458b, and HD\,189733b), the lack of observed TTVs rules out the existence of terrestrial-mass planets in or near interior, low-order mean motion resonances. Such a system architecture would have been the result of shepherding migration \citep{raymond08}. The obtained limits rule out this evolutionary scenario in these specific systems, unless some other physical process (e.g. tidal evolution) is responsible for driving the system out of resonance after the migration has stopped. If the same kind of limits are obtained for other similar systems in the future, then that would indicate that the shepherding of terrestrial-mass planets by inward migrating gas giants rarely plays a role in the evolutionary history of planetary systems.

In the case of the planet HD\,209458b, limits from the lack of TTVs and radial velocity variations together significantly constrain the existence of a perturbing planet as a cause for its ``inflated'' nature through eccentricity pumping and the subsequent dissipation of tidal energy \citep[e.g.][]{bodenheimer01}, although not all possible perturbing planets can be ruled out. For the GJ\,436 system, the absence of TTVs for the ``Hot Neptune'' planet allowed \citet{bean08} to disprove the existence of the additional 5\,M$_{\oplus}$ planet proposed by \citet{ribas08} to explain its eccentric orbit. 

I present an analysis of the transit times for CoRoT-1b\footnote{Recently the CoRoT team changed the designation of this planet from CoRoT-Exo-1b.} to search for deviations arising from perturbations from an additional planet in the system, and to place limits on the mass and orbit of such a hypothetical planet. CoRoT-1b was one of the first two planets discovered using data from the CoRoT satellite \citep{barge08}. Therefore, it presents an interesting chance to make an assessment of the real TTV sensitivity of the long, continuous sequence of space-based transit photometry from this mission. The paper is organized as follows. In \S2 I describe the CoRoT data and reduction. In \S3 I present the light curve modeling to determine transit times for each of the individual observed events. I describe the analysis of these transit times to search for variations and place limits on additional planets in \S4. I conclude in \S5 with a discussion of the results.

\section{Observations and data reduction}
CoRoT-1 was observed as part of the first CoRoT observing run of 55 days between February 2 and April 6, 2007. Details of the observations and the original analysis that definitively established the star as a host to a transiting planet were presented by \citet{barge08}. Thirty-six transit events were observed. Part-way through the observing run it was realized that these were likely transiting planet events and the sampling rate for the photometric aperture containing CoRoT-1 was changed from once per 512\,s to once every 32\,s as the usual on-board binning of 16 exposures was switched off. The first 20 transits were observed using the nominal sampling, while the last 16 were obtained in the high-frequency mode. 

I retrieved the pipeline reduced so-called ``N2'' chromatic photometric time series of CoRoT-1 from the CoRoT archive\footnote{http://idoc-corot.ias.u-psud.fr/index.jsp}. A description of the data processing steps leading to the N2 data is given by \citet{auvergne09}. From the retrieved data I extracted only the time series points flagged with a valid status and ignored those flagged as invalid (e.g. data taken while the satellite passed through the South Atlantic Anomaly or while it was entering or exiting the Earth's penumbra).

I made a few modifications to the extracted data before analyzing them to determine the transit parameters. For the data with the 512\,s sampling, the times given in the N2 data are at the end of first 32\,s exposure in the sequence of the 16 exposures that are binned together. For the data with the 32\,s sampling, the times given in the N2 data are at the end of exposure. I applied the appropriate corrections to the time stamps so that they corresponded to the midpoint of the exposures. I also converted the given heliocentric times to the reference frame of the barycenter, although this correction was relatively small (1.7\,s on average).

The chromatic N2 data contain time series obtained in three different spectral channels referred to as the blue, green, and red channels. I inspected these data for abnormalities indicative of systematic effects that were not fixed in the normal CoRoT pipeline processing. \citet{barge08} noted that their data, which were based on a preliminary reduction, were corrupted by strong cosmic ray events during two of the transits. In the version of the data that I worked with, I noticed several discontinuities attributable to cosmic ray strikes in the blue channel data when compared to the green and red channel data. I discarded some of these affected data and corrected the rest as described below. 

I identified one transit event (\#30) for which the data were too corrupted by a strong cosmic ray strike for light curve modeling, and none of the data in the date range between 43.9 and 44.8\,d after the start of the observing run were included in further analysis. The blue channel data after this were normalized to the typical level seen before the event. 

One other strong cosmic ray event was seen in the blue channel data at 1.27\,d after the beginning of the observing run. After this event, the blue channel flux exhibits an exponential decay back to the normal level over the next 18\,d. I corrected this part of the blue channel flux with a method like that used by \citet{aigrain08} to correct for a similar event in the data for CoRoT-Exo-4. The goal was to correct the blue channel data so that the blue-to-green and blue-to-red channel flux ratios were smoothly and slowly varying functions of time similar to the green-to-red channel flux ratio. To do this, I fit a power-law to the blue channel flux over the affected range. I limited the fit to data well outside of a transit. I divided the best fit from all the data in the affected range (i.e. including data during transits) with the overall normalization set by the typical flux just before the event.

After applying the described corrections to the blue channel flux, I summed the data from the three spectral channels to yield a ``white'' photometric time series. The N2 photometric counts are given as number of detected photoelectrons per second so I multiplied each sample by its effective exposure time to give the total number of counts. I took the square root of these values as the corresponding photon-limited uncertainty. The median uncertainty in the 512\,s samples was 94\,ppm, while the median uncertainty in the 32\,s samples was 375\,ppm.

As a check of the effect of the applied corrections to the blue channel on my final results, I applied the light curve modeling (see \S3) to different realizations of the data. In addition to the nominal analysis of the corrected data, I also fit a version of the data where no corrections were applied (but still ignoring the same parts of the data considered irrecoverably corrupted), and a version of the data where the blue channel data were not included in white light curve. In all cases the determined transit parameters were consistent at the level expected from their post priori uncertainties. The residuals from the fit to the corrected data were the lowest, which is mainly due to the significantly improved data for the transits immediately following the cosmic ray event that the data were corrected for. Therefore, I conclude that the corrections have the desired effect of improving the photometric data, and that this simply results in increased precision on the determined transit parameters rather than a large systematic effect on the parameters themselves.

\section{Light curve modeling}
I modeled the white light curve specified above to determine the parameters that best describe the observed transits. For each of the transits with good data, I extracted the portion of the light curve that occurred within 0.4\,d from the central transit time predicted using the ephemeris given by \citet{barge08}. This yielded 35 individual light curves. I used the exact analytic formulas including quadratic limb darkening given by \citet{mandel02} to create the model that was fitted to each of these light curves.

The global parameters of the model were the ratio of the planet and host star radii ($R_{p}/R_{\star}$),the ratio of the planet orbital semi-major axis and host star radius ($a/R_{\star}$), the planet orbital inclination ($i$), and the quadratic limb darkening coefficients ($\gamma_{1}$ and $\gamma_{2}$). I determined unique central transit times ($T_{c}$), and flux normalizations and linear trends for each of the 35 transit events. All the transit light curves were fit at the same time to simultaneously determine the global parameters and the individual event parameters. For all the modeling I assumed the transiting planet was on a circular orbit with a fixed period. I first carried out the analysis using the orbital period given by \citet{barge08}. After this, I re-determined the orbital period based on the measured individual transit times (see \S4.1) and then repeated the light curve modeling with this new period. 

I used a Levenberg-Marquardt algorithm to determine the parameters that yielded the best-fit model to the observed data. The standard $\chi^{2}$ parameter was used as the fit quality metric throughout. I applied the algorithm iteratively to reject outliers and revise the photometric error estimates. I began by first fitting the light curves assuming the photon-limited uncertainties. After the best-fit model was identified, I iteratively rejected highly deviant points and re-fit the data. The rejection threshold was set for each of the individual transit light curves to be four times the rms of the residuals around the best-fit model. This step resulted in 1.8\% of the points being eliminated.

In the next step, I calculated an adjustment factor for the photon-limited uncertainties. This factor was given by the square root of the the reduced $\chi^2$ for the best-fit to all the data together (minus the data points rejected in the previous step). The value was found to be 7.3, which indicates much larger true uncertainties in the photometry then that given by counting statistics alone. The reason for this discrepancy is unknown. I multiplied the photon limited uncertainties by this factor and then re-fit the data a final time. After the adjustment, the median uncertainty in the 512\,s samples was 681\,ppm, while the median uncertainty in the 32\,s samples was 2724\,ppm.

\begin{figure*}[ht!]
\resizebox{\hsize}{!}{\includegraphics{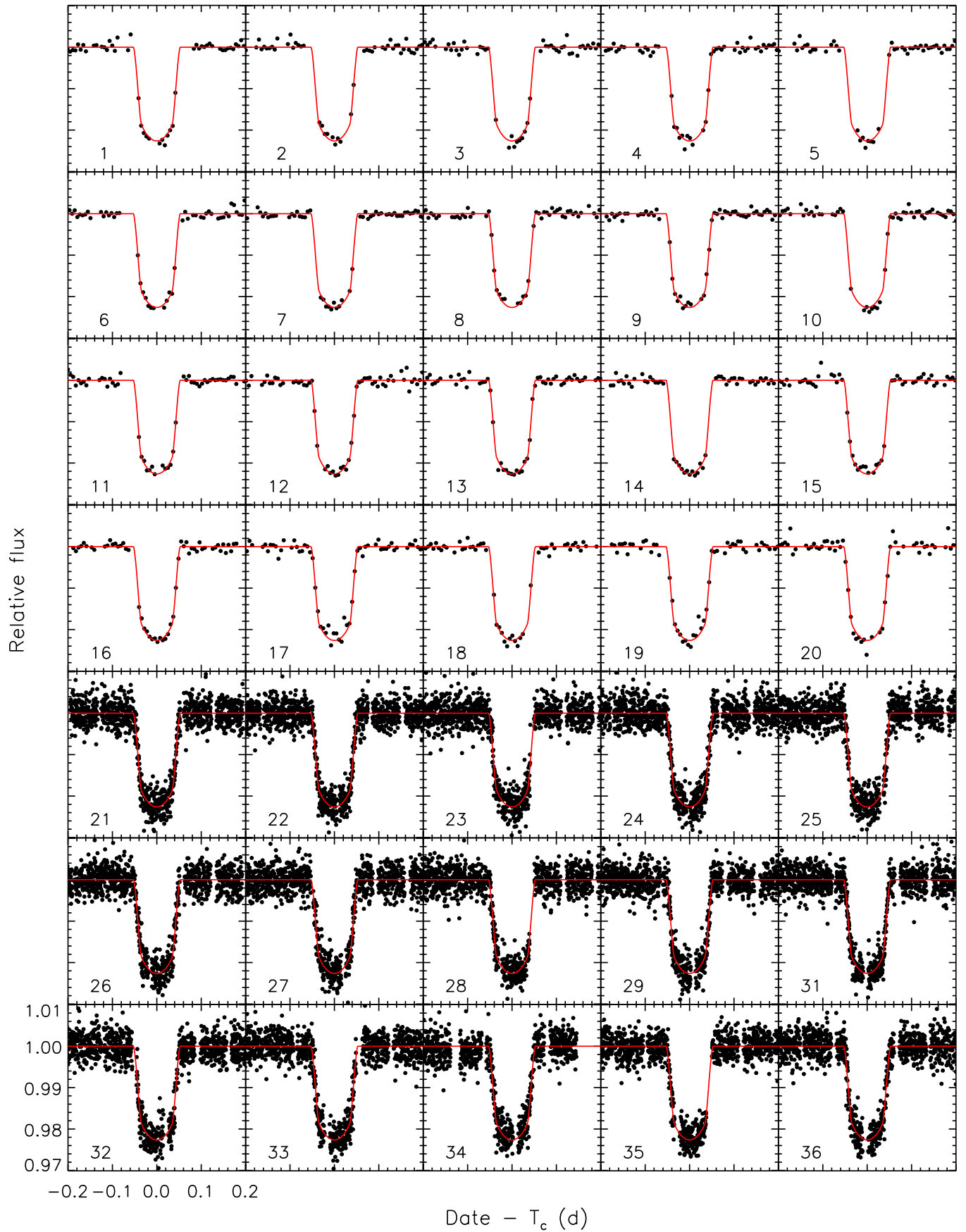}}
\caption{Individual normalized transit light curves for CoRoT-1b (points) with the best-fit model (lines). The number in each panel indicates which transit event is plotted.}
\label{f1}
\end{figure*}

The transit light curves and best-fit model are shown in Fig.~\ref{f1}. The gaps in the data are the sections of the time series that were flagged as invalid from the CoRoT pipeline. The fit residuals are Gaussian distributed, which is evidence that validates the global adjustment to the photon-limited uncertainties based on the initial reduced $\chi^2$ value.

To estimate the uncertainties in the determined parameters, I used the residual permutation boostrap or ``prayer bead'' method. I generated 10\,000 simulations of the transit light curves by adding to the best-fit model the original fit residuals shifted about a random number. I fitted each of these simulated data sets in the same way as I fit the real data. The standard deviations of the resulting parameter distributions were taken to be the parameter uncertainties. The best-fit global parameters and their corresponding uncertainties are given in Table~\ref{t1}. The transit times and uncertainties are given in Table~\ref{t2}. 

My results for the physical and orbital parameters of planet and star are slightly different than the values given by \citet{barge08}. I find a larger ratio of the planet and host star radii, a smaller ratio of the planet orbital semimajor axis and host star radius, and a lower planet orbital inclination all at about 2\,$\sigma$ formal confidence. As discussed above, these results are rather insensitive to the additional reductions I applied to the pipeline processed data. Therefore, the difference between my result and that of \citet{barge08} probably arises from differences in data themselves. The data I analyzed were processed with a more recent version of the CoRoT pipeline, whereas the data \citet{barge08} analyzed were processed with a preliminary version of the pipeline. It is likely the more recent pipline-reduced data are of superior quality due to better corrections for systematic effects that were developed as the CoRoT mission has progressed \citep{auvergne09}. All of my determined transit parameters have lower uncertainties despite my using a similar error estimation method (residual permutation boostrap) as \citet{barge08}. This suggests the more recently reduced data are indeed of better quality. I conclude that my determined transit parameters are probably also correspondingly more robust as well, and I utilize the individual transit times as described below. 

\begin{table}
\caption{Global transit parameters for CoRoT-1b.}
\label{t1}
\centering
\begin{tabular}{ll}
\hline\hline\\[-3mm]
Parameter & Value \\
\hline\\[-3mm]
$R_{p}/R_{\star}$   & $0.1433\,\pm\,0.0010$ \\
$a/R_{\star}$       & $4.751\,\pm\,0.045$ \\
$i$ ($\degr$)       & $83.88\,\pm\,0.29$ \\
$\gamma_{1}$        & $0.57\,\pm\,0.10$ \\
$\gamma_{2}$        & $-0.16\,\pm\,0.18$ \\
\hline\hline
\end{tabular}
\end{table}

\begin{table}
\caption{Transit times and residuals from the mean ephemeris for CoRoT-1b.}
\label{t2}
\centering
\begin{tabular}{cc}
\hline\hline\\[-3mm]
$T_{c}$ & O - C \\
(BJD)   & (s) \\
\hline\\[-3mm]
2454138.32761 $\pm$ 0.00047  &   21.3 \\
2454139.83712 $\pm$ 0.00059  &   68.3 \\
2454141.34485 $\pm$ 0.00062  &  -38.2 \\
2454142.85425 $\pm$ 0.00039  &   -0.7 \\
2454144.36411 $\pm$ 0.00163  &   76.6 \\
2454145.87255 $\pm$ 0.00042  &   31.5 \\
2454147.38076 $\pm$ 0.00048  &  -34.1 \\
2454148.88942 $\pm$ 0.00038  &  -60.5 \\
2454150.39899 $\pm$ 0.00026  &   -8.1 \\
2454151.90798 $\pm$ 0.00045  &   -6.4 \\
2454153.41661 $\pm$ 0.00041  &  -35.3 \\
2454154.92615 $\pm$ 0.00035  &   14.5 \\
2454156.43527 $\pm$ 0.00026  &   28.1 \\
2454157.94481 $\pm$ 0.00073  &   77.5 \\
2454159.45331 $\pm$ 0.00050  &   36.9 \\
2454160.96252 $\pm$ 0.00029  &   58.4 \\
2454162.47066 $\pm$ 0.00045  &  -13.2 \\
2454163.97926 $\pm$ 0.00044  &  -44.4 \\
2454165.48852 $\pm$ 0.00047  &  -18.9 \\
2454166.99747 $\pm$ 0.00054  &  -20.6 \\
2454168.50619 $\pm$ 0.00028  &  -41.4 \\
2454170.01596 $\pm$ 0.00020  &   27.3 \\
2454171.52372 $\pm$ 0.00028  &  -76.5 \\
2454173.03365 $\pm$ 0.00026  &    7.1 \\
2454174.54240 $\pm$ 0.00033  &  -11.6 \\
2454176.05183 $\pm$ 0.00026  &   28.2 \\
2454177.56026 $\pm$ 0.00023  &  -18.2 \\
2454179.06932 $\pm$ 0.00029  &   -9.9 \\
2454180.57844 $\pm$ 0.00025  &    3.7 \\
2454183.59609 $\pm$ 0.00024  &  -20.5 \\
2454185.10512 $\pm$ 0.00031  &  -15.3 \\
2454186.61471 $\pm$ 0.00031  &   38.4 \\
2454188.12338 $\pm$ 0.00050  &   12.8 \\
2454189.63227 $\pm$ 0.00038  &    6.3 \\
2454191.14143 $\pm$ 0.00026  &   23.5 \\
\hline\hline
\end{tabular}
\end{table}

\section{Transit time analysis}
\subsection{Search for perturbations}
\begin{figure}
\resizebox{\hsize}{!}{\includegraphics{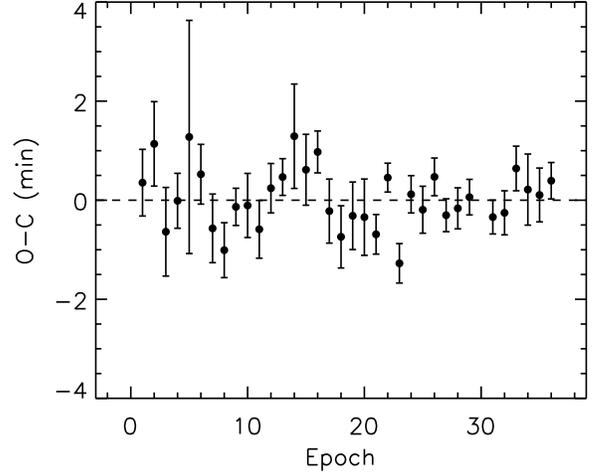}}
\caption{Transit timing residuals for CoRoT-1b.}
\label{f2}
\end{figure}

To search for TTVs, I fit the determined transit times with a model assuming a constant period and examined the residuals. The obtained mean transit time and period are given in Table~\ref{t3}. The residuals from the fit are plotted in Fig.~\ref{f2}. The rms of the residuals is 37\,s and the maximum deviation is 78\,s. The $\chi^2$ of the fit is 43.2 for 33 degrees of freedom (reduced $\chi^{2}$\,=\,1.31). The probability for a value drawn from the $\chi^2$ distribution to equal or exceed this value is 11\%.

Although the reduced $\chi^{2}$ for the fit to the transit times is somewhat higher than would be expected for a constant periodicity and well estimated errors, the significance of the discrepancy is low. Furthermore, the determined time for one transit (\#23) is essentially solely responsible for the larger than expected $\chi^2$ because it is deviant by 3.2 times its uncertainty. I closely examined the light curve for this event and found that the data did not exhibit any obvious signs of systematic error. Removing this transit time from the data set and re-fitting yielded a reduced $\chi^{2}$\,=\,1.01. Additionally, the residuals closely follow a Gaussian distribution and no residual point exceeds the standard deviation of the group by more than a factor of 2.1. Therefore, I conclude that the estimated transit time errors are reasonable, and that there is no evidence for TTVs given the precision of the data.

\begin{table}
\caption{Ephemeris for CoRoT-1b.}
\label{t3}
\centering
\begin{tabular}{lc}
\hline\hline\\[-3mm]
Parameter & Value \\
\hline\\[-3mm]
$T_{c}$ (BJD) &  2454159.452879 $\pm$ 0.000068\\
$P$ (d)       &  1.5089656 $\pm$ 0.0000060\\
\hline\hline
\end{tabular}
\end{table}

\subsection{Limits on additional planets}
As the transit times do not exhibit any evidence for perturbations to the transiting planet by another body, I turned my attention to placing limits on the mass and orbit of a hypothetical additional planet in the system. I began by first delineating the orbital parameter space such a planet could exist in based on a stability argument. To do this, I ran a long-term N-body simulation of CoRoT-1b and some massless test particles using the Mercury code \citep{chambers99}. The test particles were distributed between orbital periods of 0.1\,d and 15\,d ($a$\,=\,0.004 to 0.117\,AU) in steps of 0.02\,d. The simulation was run for 10$^{6}$ orbits of CoRoT-1b (1.5 x 10$^{6}$\,d). At the end of the simulation I determined for which orbital periods the particles did not become destabilized leading to collisions with the central star or the planet, or ejection. I found that no test particles remained stable between the central star and the planet. Outside the planet's orbit, I found that test particles remained in stable orbits for periods longer than 2.77\,d ($a$\,$>$\,0.038\,AU). 

With the region of stability for a hypothetical additional planet established, I then calculated the maximum mass such a planet could have for a given orbital period in this region and not perturb the transiting planet so much that its transit times would be inconsistent with the observed transit times. I followed the methodology used by \citet{miller-ricci08a, miller-ricci08b} for this step. The technique is based on the principle that TTVs must exhibit some non-linearity to be detectable. That is, they must be distinguished from just an incorrect assumed period for the transiting planet.

I integrated the orbits of the transiting planet and a test planet over the timespan of the observations using the Burlish-Stoer integrator in the Mercury code with the methodology described by \citet{bean08} to generate model transit times. The transiting planet's orbital parameters were initialized at their nominal values. The test planet was initialized on a circular orbit coplanar with the transiting planet. These simplifying assumptions about the orbit of the second planet are reasonable because an additional planet in a non-coplanar and/or eccentric orbit would tend to lead to even larger TTVs. The TTV signal was calculated over a grid of possible mean anomaly values for a given test planet orbital period and mass (0$\degr$ -- 360$\degr$ in steps of 1$\degr$) to marginalize over this parameter.

The transit times predicted from a given orbital integration were subtracted from the observed times to give the residuals. I fit these residuals with a first order polynomial (i.e. a linear trend). For a given period, I started with a test planet mass of 0.1\,M$_{\oplus}$ and increased this value until the smallest fit $\chi^2$ in the grid of mean anomaly values degraded by more than a certain amount from the $\chi^2$ of the best-fit constant period model to the observed transit times. I adopted limits corresponding to 3$\sigma$ confidence ($\Delta\chi^2$\,=\,9). The calculations were done for orbital periods between 2.77\,d (i.e. the shortest period for which a massless test particle was stable) and 10.0\,d in steps of 0.01\,d. A finer grid of points with steps of 0.002\,d was used around the 1:2, 3, and 4 mean motion resonances to better resolve the limits in these areas where the perturbations would be the most sensitive to the orbital configuration.

\begin{figure*}[ht!]
\resizebox{\hsize}{!}{\includegraphics{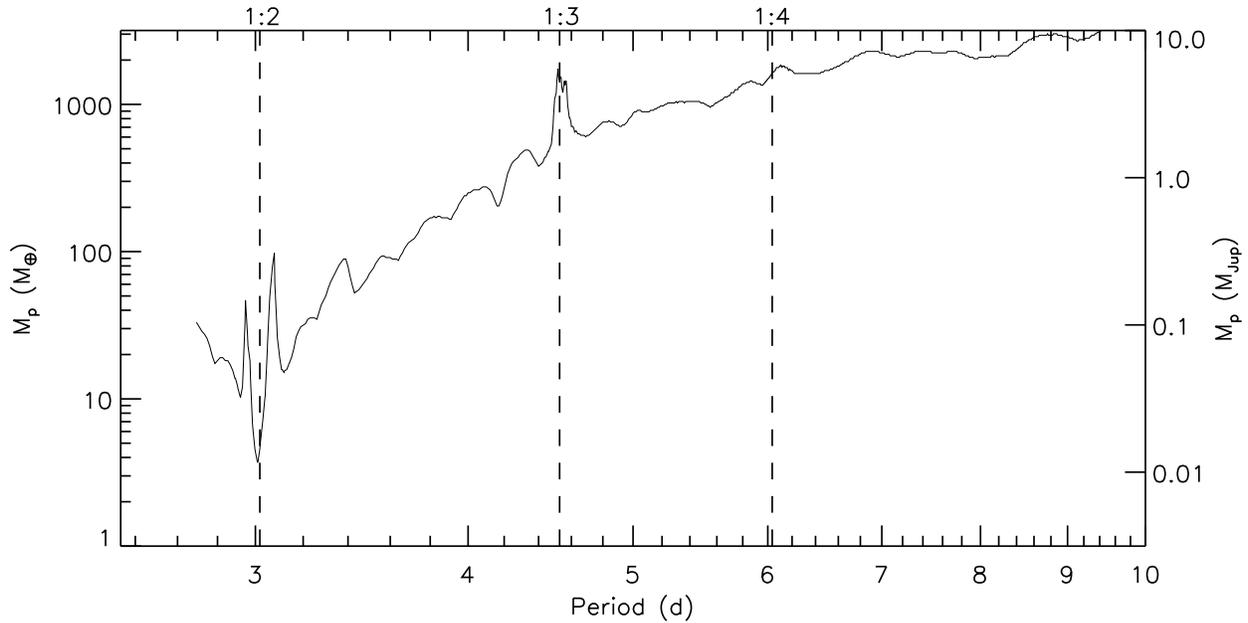}}
\caption{Upper mass limits (3\,$\sigma$ confidence) from the transit timing analysis for an additional planet in the CoRoT-1 system as a function of orbital period. The dashed lines indicate the orbital periods corresponding to the 1:2, 1:3, and 1:4 mean motion resonances with the transiting planet.}
\label{f3}
\end{figure*}

\begin{figure*}[ht!]
\resizebox{\hsize}{!}{\includegraphics{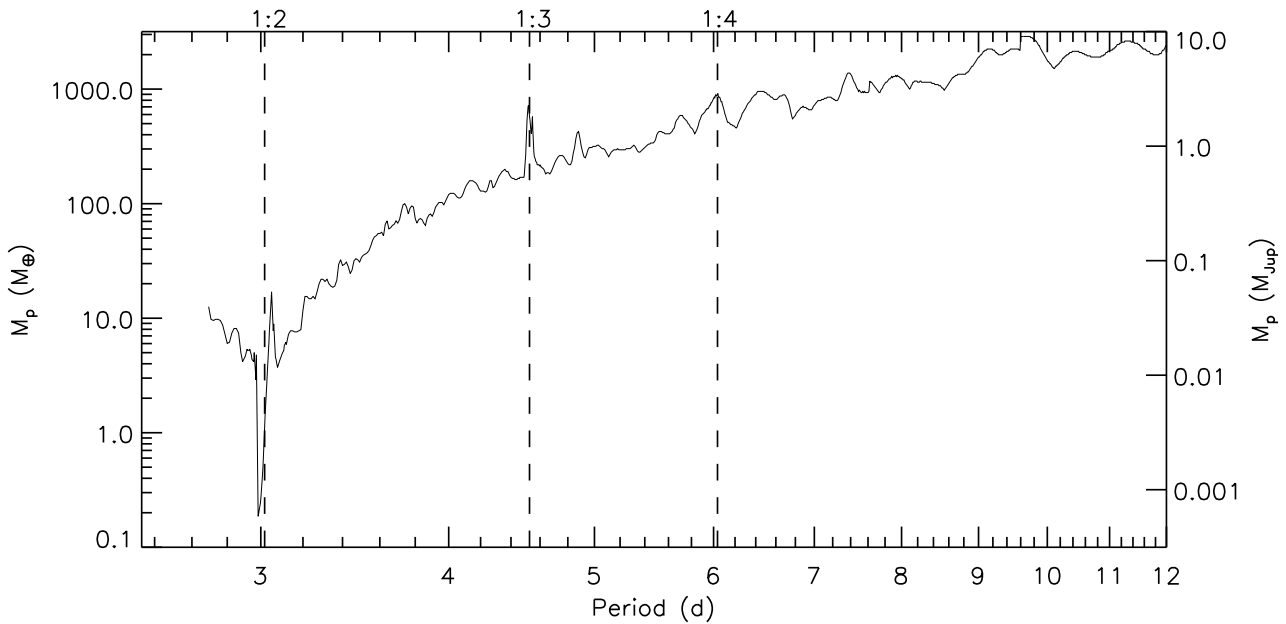}}
\caption{Same as Fig.~\ref{f3} except that the data are from an analysis of simulated transit times for a system like CoRoT-1 observed during a long run.}
\label{f4}
\end{figure*}

The results of the limit calculations are shown in Fig.~\ref{f3}. Masses greater than 4\,M$_{\oplus}$ are ruled out for a planet in a 1:2 mean motion resonance with CoRoT-1b. Interestingly, the data yield less stringent limits on planets near a 1:3 mean motion resonance ($\sim$5\,M$_{Jup}$) than in the surrounding parameter space ($\sim$2\,M$_{Jup}$). Planets with masses of 1\,M$_{Jup}$ are ruled out for all orbital periods less than about 4.2\,d. Planets with masses of 10\,M$_{Jup}$ can only be ruled out for orbital periods less than 9.4\,d on the basis of the transit times. However, the presence of such a massive planet would likely lead to instability so the true limits are likely lower than this.

\subsection{Simulation of possible CoRoT TTV sensitivity}
The data for CoRoT-1 were obtained during one of the so-called CoRoT ``short runs'' \citep{baglin06}, and for only part of the time with the high-cadence sampling. I investigated with a simulation what limits could be placed on additional planets in a similar system using transit times measured over the course of a ``long run'' of 150\,d with the high-cadence sampling the entire time. This situation represents the best possible scenario for the sensitivity of CoRoT alone to detect additional planets in systems with a transiting short-period Jovian planet. 

For the simulation I generated a sequence of transit times from 100 consecutive orbits of a planet with the same parameters of CoRoT-1b. I added to these simulated times random noise with a standard deviation of 24\,s, which is the rms of the CoRoT-1b transit times from a constant ephemeris when the high-cadence sampling was used. I then analyzed these data to determine mass limits for additional planets using the same method as above. The results are shown in Fig.~\ref{f4}.

I find that for such a scenario, the transit times would typically be about two times more sensitive for a given period. Most interestingly, the data would be sensitive to planets in and near the 1:2 mean motion resonance with masses as small as twice that of Mars (0.2\,M$_{\oplus}$). Planets with masses of 1\,M$_{Jup}$ would yield significant TTVs for all orbital periods less than about 5.4\,d. In addition, such data would be useful to probe for planets with masses of 10\,M$_{Jup}$ and orbital periods less than 12.1\,d.

\section{Summary and discussion}
I have analyzed the light curve for the transiting Jovian planet host star CoRoT-1 that was obtained with the CoRoT satellite. My results for the physical and orbital parameters of the star and the transiting planet from modeling these data are somewhat inconsistent with the results from the original analysis of the data presented by \citet{barge08}. This is most likely due to those authors analyzing a version of the light curve from a preliminary reduction of the raw data, whereas my results are based on an analysis of a version of the data produced by a more recent, and likely more robust, version of the CoRoT pipeline. The most interesting discrepancy comes for the planet-to-star radii ratio, with my results indicating a slightly larger planet than \citet{barge08}. If I assume the radius of the host star is 1.11\,$\pm$\,0.05\,M$_{\odot}$ \citep{barge08}, then my result suggests the radius of the planet is 1.54\,$\pm$\,0.07\,R$_{Jup}$. CoRoT-1b is therefore likely another ``inflated'' Hot Jupiter in the mold of HD\,209458b \citep[e.g.][]{bodenheimer03}.

The transit times determined from the light curve analysis are consistent with a constant period and, therefore, exhibit no evidence of perturbations to the transiting planet. I used this observed constancy to set limits on the mass of a hypothetical additional planet in a nearby, stable orbit. I find that the data rule out planets with masses below 4\,M$_{\oplus}$ near the 1:2 mean motion resonance, although the upper mass limits are typically much higher over the orbital period range considered. Interesting limits can only be obtained for orbital periods less than 9\,d. I confirm the general result noted in previous TTV analyses \citep[][]{steffen05, agol07, miller-ricci08a, miller-ricci08b} that this kind of study is most sensitive to planets in or near the 1:2 mean motion resonance with the transiting planet.

I have also analyzed data simulated for a similar system observed during a CoRoT long run with the high-cadence light curve sampling. The purpose of this experiment was to study what is the best possible sensitivity of CoRoT alone to detect additional planets in systems with a transiting short-period Jovian planet. As expected, such data would yield increased sensitivity to additional planets over the short run data, and planets with masses down to the Mars level near the 1:2 mean motion resonance would produce high-confidence TTV signals.

The CoRoT data yield transit time precisions on the order of a few tens of seconds, whereas the transit times determined from the highest quality light curves obtained via ground-based \citep[e.g.][]{winn09} and space-based \citep[e.g.][]{knutson07} observations are $\sim$5\,s. Nevertheless, I have demonstrated that the CoRoT data are useful for TTV studies because of their unique continuous coverage. It would be interesting to follow-up CoRoT-detected planets with high-precision ground-based observations to extend the time baseline of transit time measurements. Such a combination could yield unprecedented sensitivity to low-mass planets. 

\begin{acknowledgements}
I thank the anonymous referee and Ansgar Reiners for helpful comments on a draft of this paper. Support for this work was provided by the DFG through grants GRK 1351 and RE 1664/4-1.
\end{acknowledgements}

\bibliographystyle{aa}
\bibliography{ms}

\end{document}